\documentclass[
 aps,
 10pt,
 final,
 notitlepage,
 oneside,
 twocolumn,
 nobibnotes,
 nofootinbib,
 superscriptaddress,
 noshowpacs,
 centertags]
{revtex4}

\usepackage[dvips]{graphicx}
\usepackage{url}
\usepackage{amsmath}

\def\squareforqed{\hbox{\rlap{$\sqcap$}$\sqcup$}}

\def\sq{\ifmmode\squareforqed\else{\unskip\nobreak\hfil
\penalty50\hskip1em\null\nobreak\hfil\squareforqed
\parfillskip=0pt\finalhyphendemerits=0\endgraf}\fi}

\def\degr{\hbox{$^\circ$}}

\def\utw{\smash{\rlap{\lower5pt\hbox{$\sim$}}}}

\def\udtw{\smash{\rlap{\lower6pt\hbox{$\approx$}}}}

\def\diameter{{\ifmmode\mathchoice
{\ooalign{\hfil\hbox{$\displaystyle/$}\hfil\crcr
{\hbox{$\displaystyle\mathchar"20D$}}}}
{\ooalign{\hfil\hbox{$\textstyle/$}\hfil\crcr
{\hbox{$\textstyle\mathchar"20D$}}}}
{\ooalign{\hfil\hbox{$\scriptstyle/$}\hfil\crcr
{\hbox{$\scriptstyle\mathchar"20D$}}}}
{\ooalign{\hfil\hbox{$\scriptscriptstyle/$}\hfil\crcr
{\hbox{$\scriptscriptstyle\mathchar"20D$}}}}
\else{\ooalign{\hfil/\hfil\crcr\mathhexbox20D}}%
\fi}}

\newcommand{\aaa}{Astron. and Astrophys. }

\newcommand{\aj}{Astron.~J. }

\newcommand{\mnras}{Monthly Notices Royal Astron. Soc. }

\newcommand{\pasp}{Publ. Astron. Soc. Pacific }

\begin{document}

\keywords{Methods: data analysis -- Galaxy: kinematics and
dynamics}

%


 \title{Globular clusters: absolute proper motions and Galactic orbits}

\author{\firstname{A.~A.}~\surname{Chemel'}}
\affiliation{Lomonosov Moscow State University, Faculty of
Physics, 1, bld.2, Leninskie Gory, Moscow,  119992, Russia}

\author{\firstname{E.~V.}~\surname{Glushkova}}
\email{elena.glushkova@gmail.com}
\affiliation{Sternberg Astronomical Institute, 13, Universitetskii
prospect, Moscow, 119992, Russia}
\affiliation{Lomonosov Moscow
State University, Faculty of Physics, 1, bld.2, Leninskie Gory,
Moscow,  119992, Russia}

\author{\firstname{A.~K.}~\surname{Dambis}}
\email{dambis@yandex.ru}
\affiliation{Sternberg Astronomical
Institute, Lomonosov Moscow State University, 13, Universitetskii prospect, Moscow, 119992, Russia}

\author{\firstname{A.~S.}~\surname{Rastorguev}}
\email{alex.rastorguev@gmail.com}
\affiliation{Sternberg Astronomical Institute, 13, Universitetskii
prospect, Moscow, 119992, Russia}
\affiliation{Lomonosov Moscow
State University, Faculty of Physics, 1, bld.2, Leninskie Gory,
Moscow,  119992, Russia}

\author{\firstname{L.~N.}~\surname{Yalyalieva}}
\affiliation{Sternberg Astronomical
Institute, 13, Universitetskii prospect, Moscow, 119992, Russia}
\affiliation{Lomonosov Moscow State University, Faculty of
Physics, 1, bld.2, Leninskie Gory, Moscow,  119992, Russia}

\author{\firstname{A.~D.}~\surname{Klinichev}}
\affiliation{Lomonosov Moscow State University, Faculty of
Physics, 1, bld.2, Leninskie Gory, Moscow,  119992, Russia}

\begin{abstract}
We cross-match objects from 
several different astronomical catalogs 
to determine the absolute proper proper motions of stars within the 
30-arcmin radius fields of
115 Milky-Way globular clusters with the accuracy of 1--2~mas/yr. The proper motions
are based on  positional data  recovered from
the USNO-B1, 2MASS, URAT1, ALLWISE, UCAC5, and GAIA DR1
surveys with up to 10 positions spanning an epoch difference of up to $\sim$~65~years,
and reduced to GAIA DR1 TGAS frame using UCAC5 as the reference catalog.
Cluster members are photometrically identified by selecting horizontal- 
and red-giant branch stars on color-magnitude diagrams, and
the mean absolute proper motions of the clusters with a typical formal error
of $\sim$~0.4~mas/yr are computed by averaging the proper
motions of selected members. The inferred absolute proper motions of clusters are
combined with available radial-velocity data and 
heliocentric distance estimates to compute the
cluster orbits in terms of the Galactic potential models based on Miyamoto and Nagai disk, 
Hernquist spheroid, and modified isothermal dark-matter halo (axisymmetric model without a bar) and
the same model + rotating Ferre's bar (non-axisymmetric). Five distant clusters have 
higher-than-escape velocities, most likely due to large errors of computed 
transversal velocities, whereas the computed orbits of all other clusters remain bound to the Galaxy.
Unlike previously published results, we find the bar to affect substantially the orbits of most of the
clusters, even those at large Galactocentric distances, bringing appreciable chaotization, especially
in the portions of the orbits close to the Galactic center,
and stretching out the orbits of some of the thick-disk clusters. 

\end{abstract}

\maketitle

 \section{INTRODUCTION}
\label{intro}
Globular clusters (GCs) are potentially very important tracers of the kinematics of the Galactic bulge and 
halo from the inner parts of the Galaxy to its distant outskirts. Despite their scarcity (about 150 objects)
globular clusters have a number of important advantages over other, more populous 
halo probes (e.g. RR Lyrae-type variables and blue horizontal branch stars): the sample of 
currently catalogued GCs \citep{Harris} is much more complete than the corresponding samples of RR Lyrae 
variables and blue horizontal branch stars, and due to the group nature of GSs their inferred
parameters (e.g., heliocentric distances, heavy-element abundances and radial
velocities) are much more accurate and reliable than those of individualS stars.
Unlike the situation with the space distribution of the GC population, which is known
rather accurately, the kinematical picture of this population remains 
incomplete in that we have to rely only on its one-dimensional projection
 with only one velocity component -- the line-of-sight velocity -- 
 currently known more or less accurately for most of the clusters.
The absolute proper-motion data for GCs \citep{Casetti-Dinescu1, Casetti-Dinescu2, 
Casetti-Dinescu3,
Dambis, Dinescu1, Dinescu2, Dinescu3, Dinescu4, Cioni, Feltzing, Fritz, Kharchenko,
Kupper, Majewski, Massari, Rossi, Wang, Watkins, Zoccali}
still remains rather incomplete, not very accurate, or controversial.
In this paper we use the UCAC5\citep{UCAC5} catalog, whose reference frame is based on GAIA TGAS catalog\citep{Tgas}, and
combine it with
a number of all-sky and large-scale sky surveys (USNO-B1.0 \citep{USNOB1}, 
2MASS~\citep{TMASS}, URAT1~citep{URAT}, WISE~\citep{WISE1, WISE2},  and Gaia~\citep{Gaia}) released in the last couple of decades and
based on observations acquired in $\sim$~1950--2015 in an attempt to estimate
the absolute proper motions of stars in the GC neighborhood, identify likely cluster members, and 
determine the average proper motions  for most of the catalogued GCs. To provisionally validate
our results, we estimate the average velocity components and the 
velocity dispersion components for the subsample of metal-poor GCs ([Fe/H]~$<$~--1.0) and 
compute Galactic orbits for all clusters of our sample.

\section{DATA AND TECHNIQUE}
\label{Data}

Our aim is to compute the absolute proper motions of stars in the fields of globular
clusters, identify the likely cluster members, and compute the mean proper motions
of the clusters studied by averaging the inferred proper motions for selected cluster 
members. 

We begin by estimating the proper motions of most of the stars in the fields of cluster considered
based on
the star positions recovered from a number of the most extensive sky surveys
(USNO-B1.0, 2MASS, WISE, URAT1, UCAC5, GAIA DR1) containing good positional data. 

The first step is to 
cross-match stars from these surveys in the cluster fields. To facilitate this task, we developed 
a  command-line CROSSMATCH program \citep{Klinichev} written in java programming language and serving as a convenient interface to
the well-known STILTS program~\citep{Stilts}. CROSSMATCH code is available at
\newline
{\footnotesize
\verb|www.sai.msu.ru/groups/cluster/cl/crossmatch/Crossmatch_4.3.0.zip|
}
\newline
For each cluster we begin our analysis by cross-matching stars from the above catalogs
within 30~acrmin of the cluster centers with a cross-match radius of 1 or 2~arcsec. 

The next step is to bring the positions adopted from the survey catalogs to the frame defined by
Gaia DR1 positions and proper motions. The problem is that although bona fide 2015.0 positions are
currently available for more than 1 billion Gaia stars, Gaia proper motions are available only for
$>$~2~million stars of the Gaia TGAS subset~\citep{Tgas}, which is evidently insufficient for proper
astrometric reduction in cluster fields for two reasons:
\begin{enumerate}
	\item The subset is insufficiently dense to provide enough stars per field
        \item The subset consists of too bright stars, whose images are too saturated in the surveys 
considered and therefore their measured positions are fraught with systematic errors and cannot be used to determine reduction parameters for fainter stars.   
\end{enumerate}
To overcome this difficulty, we use the much deeper and more extensive UCAC5 catalogue~\citep{UCAC5} as
our reference set instead of Gaia FR1 TGAS. According to its authors, UCAC5 should be a good extension of Gaia DR1 TGAS
down to limiting magnitudes of $\sim$~16.0$^m$ and fainter. We therefore adopt UCAC5 positions as is without applying any
corrections. Naturally, we treat  published Gaia 2015.0 positions in the same way. We bring the WISE MJD=55400.0 (2010.5589) positions
to UCAC5 frame via usual linear plate adjustment and do the same with 2MASS and URAT1 positions. As for USNO-B10,
we reconstruct the positions of the star's images on individual
Schmidt plates based on the information provided in the USNO-B1.0
Catalog \citep{USNOB1} (the J2000.0 position, proper motion,
and tangent-plane offsets
--- $B1_{\xi}$, $B2_{\xi}$, $R1_{\xi}$, $R2_{\xi}$, $I_{\xi}$ and
$B1_{\eta}$, $B2_{\eta}$, $R1_{\eta}$, $R2_{\eta}$, $I_{\eta}$
--- along the x- and y-directions with respect to the mean-epoch
position). Again, we perform linear plate adjustment
using UCAC5 stars located within 30~arcmin of the cluster
centre as reference objects and adopting
individual USNO-B1.0 plate epochs from plate logs at the page
\newline
http://www.nofs.navy.mil/data/fchpix/cfhelp.html\#plogs
\newline
of the site of USNO Flagstaff Station Integrated Image and
Catalogue Archive Service (see \citep{Dambis09}). As a result, we obtain a maximum of ten positions (a maximum of 
five reconstructed Schmidt plate 
positions  from USNO-B1.0 + Gaia position + one UCAC5 position+ one reconstructed position from each
of the 2MASS, URAT1,  and  ALLWISE catalogs
per star north of 
$\delta$=~-33.0$^o$, and a maximum of 
three reconstructed Schmidt plate 
positions  from USNO-B1.0 + Gaia position + one UCAC5 position+ one reconstructed position from each
of the 2MASS and  ALLWISE catalogs
per star south of $\delta$=~-33.0$^o$). The total epoch span usually varies from about 65 years for stars north of 
$\delta$=~-33.0$^o$ to about 30--35 years for stars south of $\delta$=~-33.0$^o$.

\begin{figure}
\includegraphics[width=0.9\linewidth]{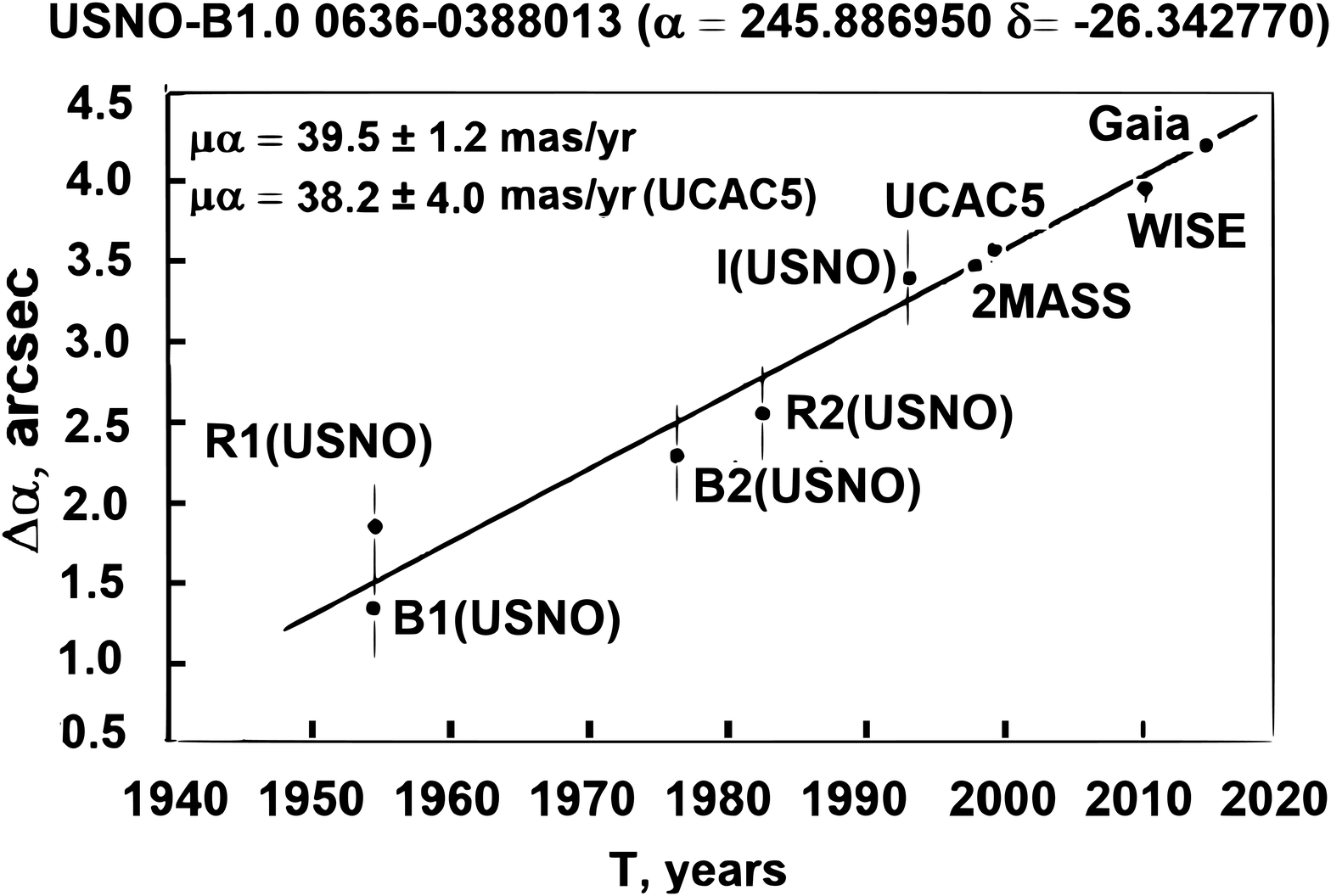}
 \caption{Right ascension difference for star USNO-B1.0~0636-0388013
as a function of epoch. The dots with errorbars show the reconstructed positions in the UCAC5 reference frame.
The solid line shows the
linear least squares fit used to determine the proper motion.}
 \label{fig:pm_alpha}
\end{figure}

\begin{figure}[]
\includegraphics[width=\linewidth]{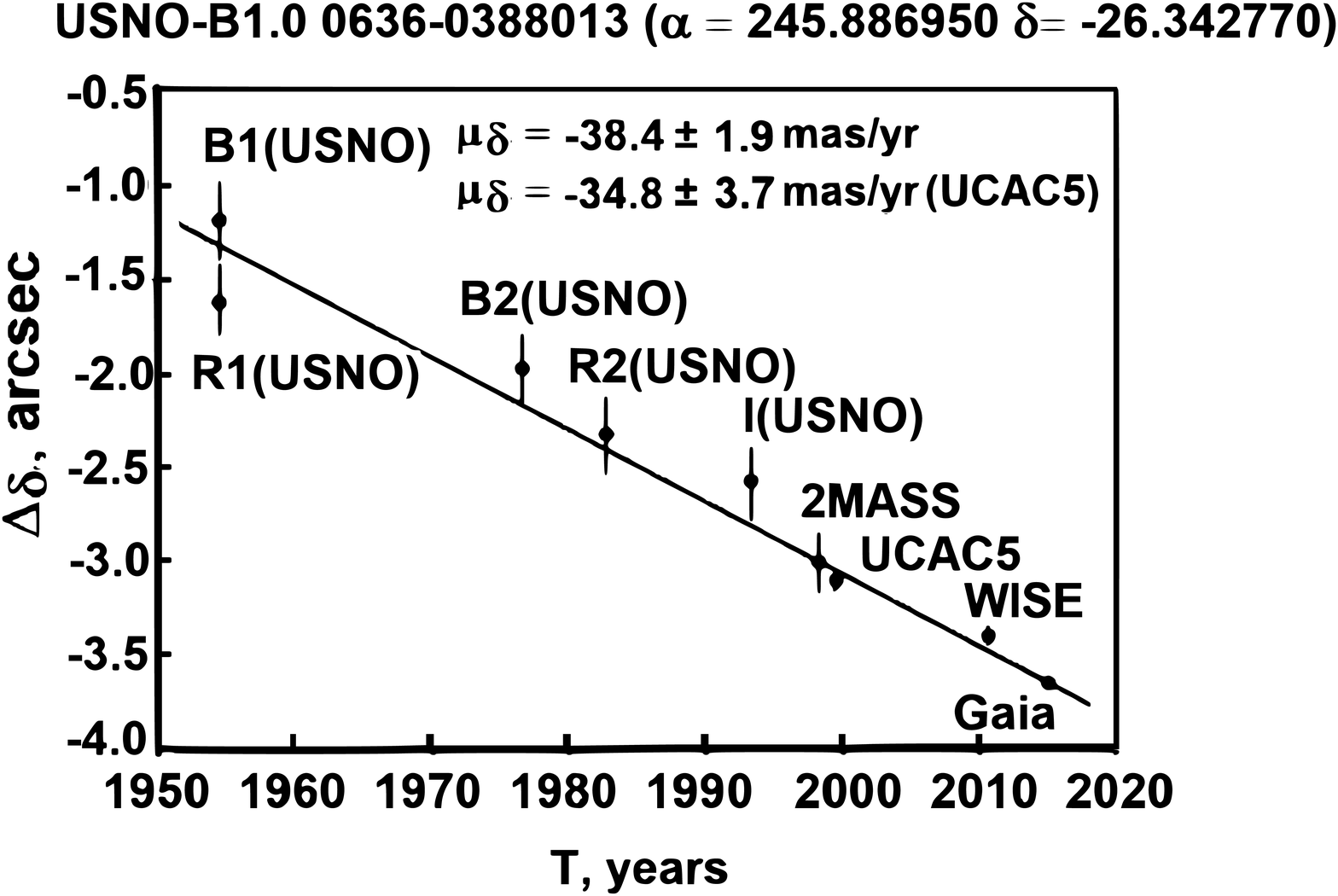}
\caption{Declination difference for star USNO-B1.0~0636-0388013
as a function of epoch. The dots with errorbars show the reconstructed positions in the UCAC5 reference frame.
The solid line shows the
linear least squares fit used to determine the proper motion.} \label{fig:pm_delta}
\end{figure}

To illustrate the procedure, we show in 
Figs.~\ref{fig:pm_alpha} and~\ref{fig:pm_delta} the variation of the right ascension and declination,
respectively, of the star USNO-B1.0~0636-0388013 as a function of epoch. The dots with the errorbars 
show the corresponding reconstructed positions and the solid lines, the linear least squares fit used
to determine the proper motion. As is evident from the figures, the use of extra positional data
improves appreciably the accuracy of the proper-motion components  compared to UCAC5.

We then identify the likely cluster members by selecting stars 
from the horizontal and red-giant  branches and the main sequence in the cluster color-magnitude diagrams
based on 2MASS data and compute the average proper-motion components along right ascension and declination.

To validate our technique, we also determine the proper motions of nearby globular clusters (i.e., those whose
horizontal- and red-giant branch stars are well within the UCAC5 limiting magnitude) based on the
original UCAC5 proper motions of the cluster members, and compare the results with those obtained for fainter, 
main-sequence stars
of the cluster determined from the UCAC5-calibrated USNO-B1.0, 2MASS, URAT1, and  WISE survey positions 
combined with UCAC5 and GAIA DR1 positions (where available). We find the bright-star and faint-star based
proper motions for nearby clusters to be fairly consistent, thereby confirming the validity of the adopted procedure 
and showing it to introduce no appreciable magnitude-dependent biases.

\section{RESULTS}
\label{Results}

\subsection{PROPER MOTIONS AND THEIR ERRORS}
\label{pmvt}

\begin{figure}[]
\includegraphics[width=\linewidth]{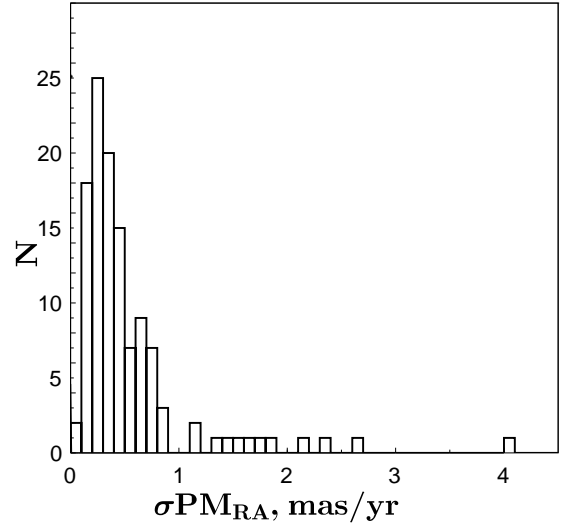}
\caption{Distribution of globular-cluster proper-motion errors
in right ascension, $\sigma$~PM$_{RA}$.} \label{fig:errpm_alpha}
\end{figure}

\begin{figure}[]
\includegraphics[width=\linewidth]{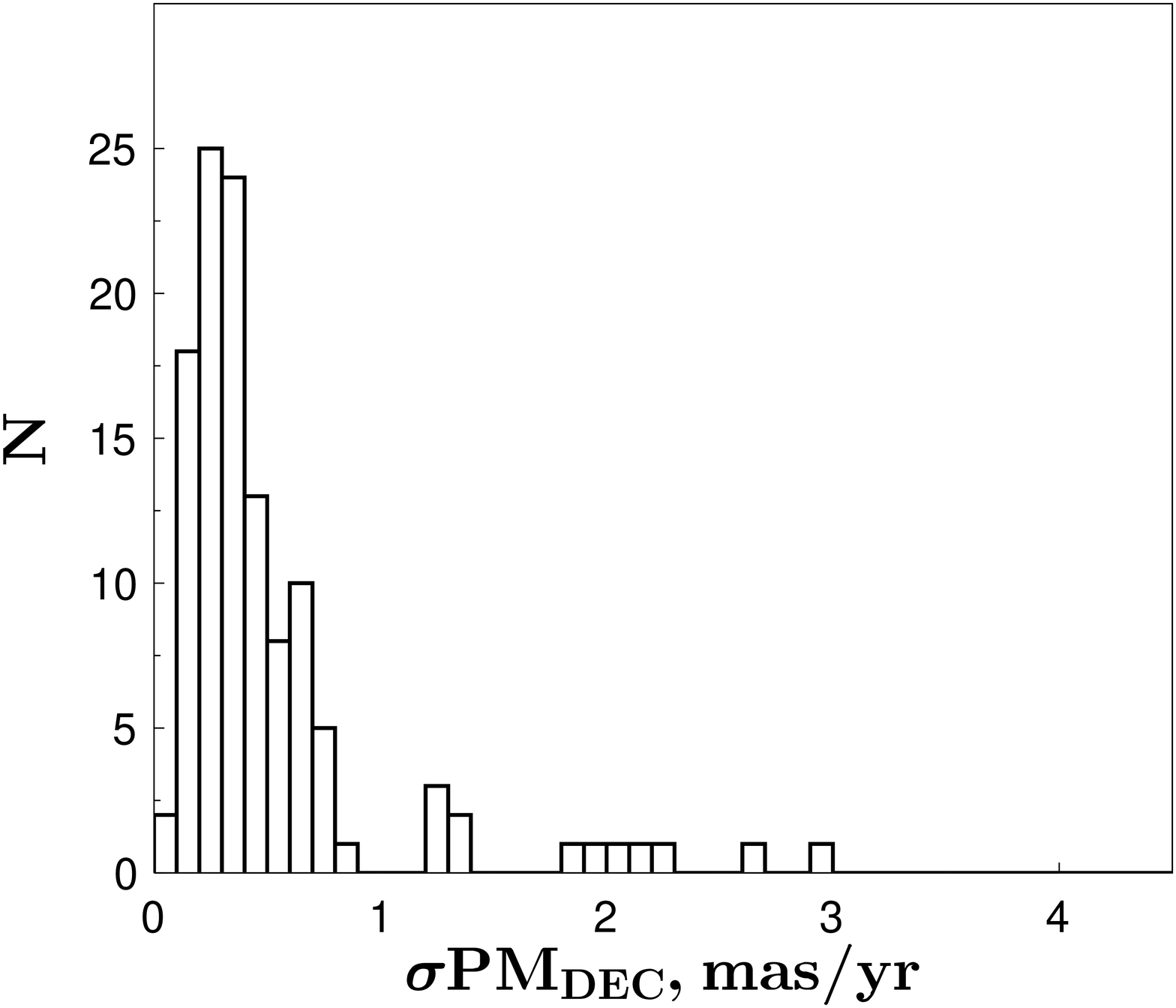}
\caption{Distribution of globular-cluster proper-motion errors
in right ascension, $\sigma$~PM$_{RA}$} \label{fig:errpm_delta}
\end{figure}

\begin{figure}[]
\includegraphics[width=\linewidth]{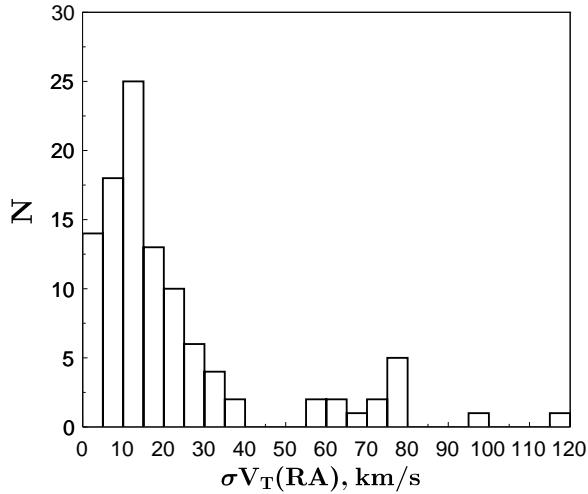}
\caption{Distribution of transversal-velocity errors $\sigma_T$~(RA) in right ascension
 for 106 globular clusters with $\sigma_PM_{RA} \leq$~1~mas/yr and
$\sigma_PM_{DEC} \leq$~1~mas/yr.} \label{fig:errvt_alpha}
\end{figure}

\begin{figure}[]
\includegraphics[width=\linewidth]{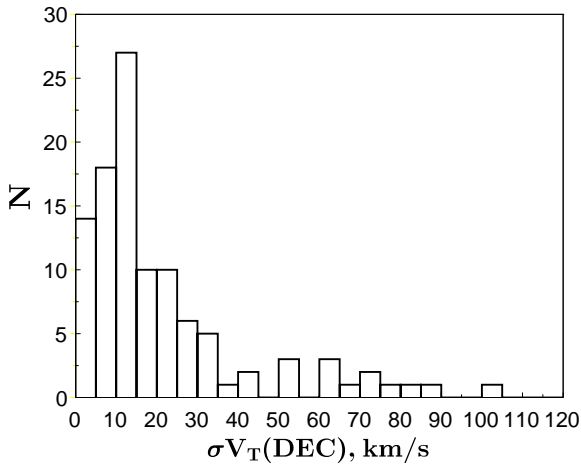}
\caption{Distribution of transversal-velocity errors $\sigma_T$~(DEC) in declination for 
106 globular clusters with $\sigma_PM_{RA} \leq$~1~mas/yr and
$\sigma_PM_{DEC} \leq$~1~mas/yr.} \label{fig:errvt_delta}
\end{figure}

\begin{figure}[]
\includegraphics[width=\linewidth]{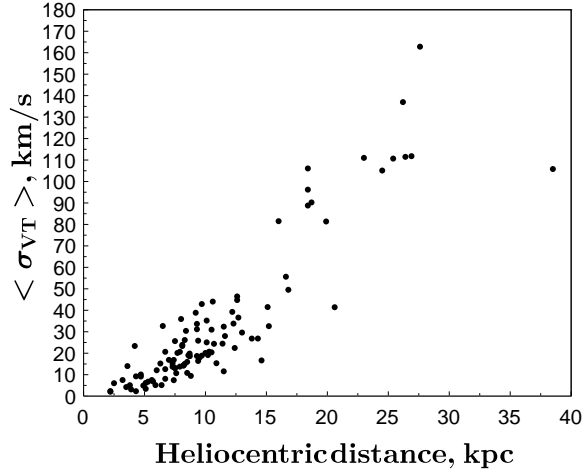}
\caption{Dependence of the transversal-velocity error 
$<\sigma_{VT}>~=~(\sigma V_T~(RA)^2~+~\sigma V_T~(DEC)^2)^{1/2}$ on heliocentric distance for 
106 globular clusters with $\sigma_PM_{RA} \leq$~1~mas/yr and
$\sigma_PM_{DEC} \leq$~1~mas/yr.} \label{fig:errvt_rhel}
\end{figure}

We used the data and technique described above to compute the absolute proper motions of stars in the fields
of 115 Milky-Way globular clusters and determine the average absolute proper motions of the clusters. The results are 
summarized in Table~\ref{gctable}. Columns~1 and 2 of give the J2000.0 equatorial coordinates of the cluster center,
column~3 and 4 give the name and alternative name of the cluster, respectively. Column~5 gives the
apparent tidal radius of the cluster in acrmin, columns 6 and 7, the apparent $V$-band magnitude of the
horizontal branch and the apparent $V$-band distance modulus, respectively. Columns 8 and 9 give the 
inferred cluster proper-motion component in right ascension and its standard error in mas/yr, respectively, and columns
10 and 11, the proper-motion component in declination and its standard error, respectively.
Columns 12 and 13 give the Galactic coordinates the cluster center,
and columns 14, 15, 16, and 17 give the average heliocentric radial velocity, heliocentric distance,
$E_{B--V}$ color excess, and metallicity [Fe/H], respectively. Practically all 
data except the proper motions (columns 8--11)
are adopted from the catalog of Harris~\citep{Harris}. The radial velocities of E3, ESO452-SC11,
and Djorg~2 are adopted from the papers \citep{delafuente}, \citep{Koch}, and \citep{Dias},
respectively. Figs.~\ref{fig:errpm_alpha} and~\ref{fig:errpm_delta} show the distributions
of the formal errors of inferred cluster proper motions in right ascension and declination, respectively. The median
errors are equal to 0.36 and 0.35 mas/yr for the proper motion in right ascension and declination, 
respectively. Figs.~\ref{fig:errvt_alpha} and~\ref{fig:errvt_delta}
show the corresponding transversal-velocity errors, $\sigma V_T$~(RA) and $\sigma V_T$~(DEC), whose median values are equal to
17 and 16 km/s, respectively. Figure~\ref{fig:errvt_rhel}  shows 
the dependence of the transversal-velocity error $<\sigma_{VT}>~=~(\sigma V_T~(RA)^2~+~\sigma V_T~(DEC)^2)^{1/2}$ on heliocentric distance.

\begin{table*}[]
\scriptsize
\caption{Globular cluster data}
\label{Data}
\medskip
\begin{tabular}{r r l l r r r r r r r r r r r r r}
  \hline
RA2000 & DEC2000 & Name &   Alternative &    $R_{tidal}$ &    V(HB) & DM &     pmra &   $e_{pmra}$ & pmde & $e_{pmde}$ & l &  b &  $V_r$ &  $D_{hel}$  &  $E_{B-V}$ & [Fe/H] \\
 & & & name & arcmin & & &  \multicolumn{4}{c}{mas/yr} & & & km/s& kpc & & \\
265.1721  & -53.6736  & NGC 6397  &              &      15.811   &    12.87  & 12.31  &     3.34  &    0.18   &     -18.00  &    0.14   &    338.17  & -11.96   &    18.9   &  2.2   &   0.18  &  -1.95  \\
245.8979  & -26.5253  & NGC 6121  &  M 4         &      32.291   &    13.45  & 12.78  &   -12.64  &    0.21   &     -18.26  &    0.19   &    350.97  &  15.97   &    70.2   &  2.2   &   0.36  &  -1.20  \\
287.7158  & -59.9819  & NGC 6752  &              &      53.759   &    13.70  & 13.08  &    -2.74  &    0.13   &      -4.40  &    0.13   &    336.50  & -25.63   &   -24.5   &  3.9   &   0.04  &  -1.55  \\
006.0217  & -72.0808  & NGC 104   &  47 Tuc      &      40.570   &    14.06  & 13.32  &     5.41  &    0.20   &      -2.79  &    0.12   &    305.90  & -44.89   &   -18.7   &  4.3   &   0.05  &  -0.76  \\
279.1008  & -23.9033  & NGC 6656  &  M 22        &      28.993   &    14.15  & 13.55  &     9.36  &    0.35   &      -6.62  &    0.34   &      9.89  &  -7.55   &  -148.9   &  3.2   &   0.34  &  -1.64  \\
294.9975  & -30.9622  & NGC 6809  &  M 55        &      16.285   &    14.40  & 13.82  &    -3.89  &    0.23   &      -9.18  &    0.22   &      8.80  & -23.27   &   174.8   &  5.3   &   0.07  &  -1.81  \\
298.4421  & +18.7783  & NGC 6838  &  M 71        &       8.899   &    14.44  & 13.70  &    -5.46  &    0.20   &      -2.55  &    0.20   &     56.74  &  -4.56   &   -22.9   &  3.8   &   0.25  &  -0.73  \\
201.6912  & -47.4769  & NGC 5139  & $\omega$ Cen &      44.835   &    14.53  & 13.92  &    -3.37  &    0.10   &      -6.84  &    0.10   &    309.10  &  14.97   &   232.3   &  5.1   &   0.12  &  -1.62  \\
251.8104  &  -1.9478  & NGC 6218  &  M 12        &      15.832   &    14.60  & 13.97  &     0.24  &    0.84   &      -6.98  &    0.74   &     15.72  &  26.31   &   -42.1   &  4.7   &   0.19  &  -1.48  \\
254.2871  &  -4.0994  & NGC 6254  &  M 10        &      21.602   &    14.65  & 14.03  &    -4.83  &    0.37   &      -6.15  &    0.29   &     15.14  &  23.08   &    75.8   &  4.3   &   0.28  &  -1.52  \\
154.4033  & -46.4111  & NGC 3201  &              &      29.605   &    14.80  & 14.17  &     9.92  &    0.19   &      -2.90  &    0.35   &    277.23  &   8.64   &   494.0   &  5.1   &   0.21  &  -1.48  \\
140.2471  & -77.2825  & E 3       &              &      10.516   &    14.80  & 14.07  &    -5.56  &    0.55   &       2.28  &    0.55   &    292.27  & -19.02   &    45.0   &  4.2   &   0.30  &  -0.80  \\
271.8358  & -24.9975  & NGC 6544  &              &       2.133   &    14.90  & 14.28  &    -2.23  &    0.37   &     -17.43  &    0.45   &      5.84  &  -2.20   &   -27.3   &  2.5   &   0.74  &  -1.56  \\
250.4229  & +36.4603  & NGC 6205  &  M 13        &      27.195   &    14.90  & 14.28  &    -3.05  &    0.48   &      -1.15  &    0.48   &     59.01  &  40.91   &  -246.6   &  7.0   &   0.02  &  -1.54  \\
229.6408  &  +2.0828  & NGC 5904  &  M 5         &      29.652   &    15.07  & 14.41  &     4.01  &    0.28   &      -6.25  &    0.34   &      3.86  &  46.80   &    52.1   &  7.3   &   0.03  &  -1.29  \\
259.2804  & +43.1364  & NGC 6341  &  M 92        &      14.850   &    15.10  & 14.59  &    -4.16  &    0.43   &      -0.56  &    0.51   &     68.34  &  34.86   &  -121.6   &  8.1   &   0.02  &  -2.29  \\
325.0917  & -23.1792  & NGC 7099  &  M 30        &      18.974   &    15.10  & 14.57  &     0.71  &    0.40   &      -7.28  &    0.40   &     27.18  & -46.83   &  -184.2   &  7.9   &   0.03  &  -2.12  \\
261.3717  & -48.4228  & NGC 6352  &              &      10.449   &    15.13  & 14.39  &    -2.15  &    0.21   &      -4.85  &    0.21   &    341.42  &  -7.17   &  -120.9   &  5.6   &   0.21  &  -0.70  \\
272.0092  & -43.7056  & NGC 6541  &              &      30.000   &    15.30  & 14.72  &    -1.36  &    0.34   &      -6.90  &    0.36   &    349.29  & -11.18   &  -156.2   &  7.4   &   0.12  &  -1.83  \\
271.5358  & -27.7653  & NGC 6540  &  Djorg 3     &       9.487   &    15.30  & 14.60  &    -2.89  &    0.33   &      -4.88  &    0.34   &      3.29  &  -3.31   &   -17.7   &  3.5   &   0.60  &  -1.00  \\
186.4392  & -72.6592  & NGC 4372  &              &      34.917   &    15.30  & 14.76  &    -6.45  &    0.23   &       2.60  &    0.30   &    300.99  &  -9.88   &    72.3   &  4.9   &   0.42  &  -2.09  \\
013.1979  & -26.5900  & NGC 288   &              &      12.951   &    15.30  & 14.64  &     3.49  &    0.69   &      -5.33  &    0.44   &    152.28  & -89.38   &   -46.6   &  8.1   &   0.03  &  -1.24  \\
262.9783  & -67.0481  & NGC 6362  &              &      16.618   &    15.34  & 14.65  &    -5.62  &    0.31   &      -5.02  &    0.26   &    325.55  & -17.57   &   -13.1   &  7.5   &   0.09  &  -1.06  \\
015.8096  & -70.8483  & NGC 362   &              &      14.806   &    15.43  & 14.75  &     6.47  &    0.48   &      -1.55  &    0.49   &    301.53  & -46.25   &   223.5   &  8.3   &   0.05  &  -1.16  \\
194.8958  & -70.8747  & NGC 4833  &              &      17.783   &    15.45  & 14.87  &    -7.78  &    0.14   &      -1.79  &    0.18   &    303.61  &  -8.01   &   200.2   &  5.9   &   0.33  &  -1.79  \\
284.8883  & -36.6317  & NGC 6723  &              &      10.547   &    15.50  & 14.82  &     1.20  &    0.34   &      -2.80  &    0.35   &      0.07  & -17.30   &   -94.5   &  8.6   &   0.05  &  -1.12  \\
280.8029  & -32.2919  & NGC 6681  &  M 70        &       9.487   &    15.55  & 14.93  &     1.14  &    0.33   &      -5.70  &    0.33   &      2.85  & -12.51   &   218.6   &  8.7   &   0.07  &  -1.51  \\
283.7758  & -22.7008  & NGC 6717  &  Pal 9       &       9.399   &    15.56  & 14.90  &    -4.52  &    1.16   &      -5.90  &    2.08   &     12.88  & -10.90   &    22.8   &  7.1   &   0.21  &  -1.29  \\
261.9346  &  -5.0767  & NGC 6366  &              &      15.221   &    15.65  & 14.92  &    -0.02  &    0.58   &      -5.52  &    0.61   &     18.41  &  16.04   &  -122.3   &  3.6   &   0.69  &  -0.82  \\
205.5467  & +28.3756  & NGC 5272  &  M 3         &      35.397   &    15.65  & 15.04  &    -0.23  &    0.30   &      -3.45  &    0.29   &     42.21  &  78.71   &  -148.6   & 10.0   &   0.01  &  -1.57  \\
189.8667  & -26.7428  & NGC 4590  &  M 68        &      30.120   &    15.68  & 15.14  &    -1.52  &    0.52   &       1.17  &    0.48   &    299.63  &  36.05   &   -95.2   & 10.1   &   0.04  &  -2.06  \\
248.1329  & -13.0536  & NGC 6171  &  M 107       &      17.474   &    15.70  & 15.01  &    -2.49  &    0.37   &      -5.83  &    0.34   &      3.37  &  23.01   &   -33.8   &  6.3   &   0.33  &  -1.04  \\
276.1371  & -24.8700  & NGC 6626  &  M 28        &      11.226   &    15.70  & 15.07  &    -0.24  &    0.19   &      -7.97  &    0.23   &      7.80  &  -5.58   &    17.0   &  5.7   &   0.41  &  -1.45  \\
270.9612  &  -0.2969  & NGC 6535  &              &       8.380   &    15.73  & 15.15  &    -6.35  &    5.07   &       3.70  &    5.47   &     27.18  &  10.44   &  -215.1   &  6.8   &   0.32  &  -1.80  \\
322.4929  & +12.1669  & NGC 7078  &  M 15        &      22.136   &    15.83  & 15.31  &     2.04  &    0.46   &      -2.15  &    0.73   &     65.01  & -27.31   &  -107.5   & 10.2   &   0.09  &  -2.22  \\
278.9404  & -32.9903  & NGC 6652  &              &       4.417   &    15.85  & 15.14  &    -2.39  &    0.41   &      -5.14  &    0.39   &      1.53  & -11.38   &  -111.7   &  9.4   &   0.09  &  -0.96  \\
277.8467  & -32.3481  & NGC 6637  &  M 69        &       8.346   &    15.85  & 15.11  &    -4.48  &    0.27   &      -6.10  &    0.27   &      1.72  & -10.27   &    39.9   &  8.2   &   0.17  &  -0.71  \\
244.2604  & -22.9750  & NGC 6093  &  M 80        &      13.369   &    15.86  & 15.25  &    -2.58  &    0.33   &      -5.96  &    0.31   &    352.67  &  19.46   &     9.3   &  8.7   &   0.18  &  -1.62  \\
272.5767  & -31.7636  & NGC 6558  &              &       9.487   &    15.97  & 15.34  &    -1.38  &    0.12   &      -5.50  &    0.20   &      0.20  &  -6.03   &  -143.7   &  6.4   &   0.42  &  -1.44  \\
323.3721  &  -0.8231  & NGC 7089  &  M 2         &      21.453   &    16.05  & 15.44  &     1.22  &    0.33   &      -2.87  &    0.40   &     53.38  & -35.78   &    -5.3   & 11.4   &   0.05  &  -1.62  \\
102.2467  & -36.0053  & NGC 2298  &              &       6.479   &    16.11  & 15.54  &     3.25  &    0.62   &      -2.85  &    0.64   &    245.63  & -16.01   &   148.9   & 10.6   &   0.13  &  -1.85  \\
078.5262  & -40.0472  & NGC 1851  &              &      13.902   &    16.15  & 15.49  &     1.72  &    0.48   &      -0.10  &    0.44   &    244.51  & -35.04   &   320.9   & 12.2   &   0.02  &  -1.26  \\
081.0442  & -24.5242  & NGC 1904  &  M 79        &       8.397   &    16.15  & 15.53  &     2.68  &    0.56   &      -1.43  &    0.45   &    227.23  & -29.35   &   207.5   & 12.6   &   0.01  &  -1.54  \\
289.1479  & +30.1847  & NGC 6779  &  M 56        &       8.674   &    16.16  & 15.60  &    -2.10  &    2.68   &       2.33  &    2.64   &     62.66  &   8.34   &  -135.7   &  9.9   &   0.20  &  -1.94  \\
138.0108  & -64.8631  & NGC 2808  &              &      15.310   &    16.19  & 15.55  &     0.97  &    0.30   &      -0.40  &    0.29   &    282.19  & -11.25   &    93.6   &  9.3   &   0.23  &  -1.37  \\
283.2679  &  -8.7061  & NGC 6712  &              &       7.467   &    16.25  & 15.55  &     3.16  &    0.29   &      -5.07  &    0.28   &     25.35  &  -4.32   &  -107.7   &  6.7   &   0.46  &  -1.01  \\
255.3025  & -30.1122  & NGC 6266  &  M 62        &       9.021   &    16.25  & 15.59  &    -4.50  &    0.19   &      -2.46  &    0.18   &    353.58  &   7.32   &   -65.8   &  6.7   &   0.47  &  -1.29  \\
258.6354  & -29.4622  & NGC 6304  &              &      13.250   &    16.25  & 15.49  &    -2.71  &    0.30   &      -1.90  &    0.28   &    355.83  &   5.38   &  -107.3   &  6.0   &   0.52  &  -0.59  \\
259.7992  & -18.5164  & NGC 6333  &  M 9         &       8.193   &    16.30  & 15.71  &    -2.69  &    0.27   &      -4.22  &    0.26   &      5.54  &  10.70   &   229.1   &  8.3   &   0.36  &  -1.72  \\
277.9762  & -23.4764  & NGC 6642  &              &       9.772   &    16.30  & 15.65  &     0.33  &    0.21   &      -4.32  &    0.18   &      9.81  &  -6.44   &   -57.2   &  7.6   &   0.40  &  -1.35  \\
229.3521  & -21.0103  & NGC 5897  &              &      12.085   &    16.35  & 15.77  &    -6.36  &    0.43   &      -3.97  &    0.46   &    342.95  &  30.29   &   101.5   & 12.7   &   0.08  &  -1.80  \\
255.6571  & -26.2681  & NGC 6273  &  M 19        &      14.570   &    16.40  & 15.80  &    -4.27  &    0.21   &       1.04  &    0.21   &    356.87  &   9.38   &   135.0   &  8.5   &   0.37  &  -1.68  \\
269.7583  & -44.2650  & NGC 6496  &              &       5.262   &    16.47  & 15.72  &    -2.62  &    0.37   &      -9.14  &    0.41   &    348.02  & -10.01   &  -112.7   & 11.6   &   0.13  &  -0.64  \\
257.5433  & -26.5817  & NGC 6293  &              &      15.811   &    16.50  & 15.94  &     0.65  &    0.16   &      -3.97  &    0.16   &    357.62  &   7.83   &   -98.9   &  8.8   &   0.39  &  -1.92  \\
236.5146  & -37.7861  & NGC 5986  &              &      10.455   &    16.50  & 15.90  &    -3.61  &    0.32   &      -4.66  &    0.30   &    337.02  &  13.27   &    88.9   & 10.3   &   0.27  &  -1.67  \\
277.7342  & -25.4964  & NGC 6638  &              &       6.531   &    16.50  & 15.80  &    -2.64  &    0.29   &      -3.18  &    0.29   &      7.90  &  -7.15   &    18.1   &  8.2   &   0.40  &  -0.99  \\
206.6104  & -51.3733  & NGC 5286  &              &       8.364   &    16.50  & 15.90  &    -0.25  &    0.35   &       0.48  &    0.33   &    311.61  &  10.57   &    58.3   & 10.7   &   0.24  &  -1.67  \\
274.6571  & -52.2150  & NGC 6584  &              &       9.351   &    16.53  & 15.90  &     0.70  &    0.34   &      -6.58  &    0.35   &    342.14  & -16.41   &   222.9   & 13.0   &   0.11  &  -1.49  \\
246.4525  & -72.2017  & NGC 6101  &              &       7.256   &    16.60  & 16.02  &     2.25  &    0.42   &       0.78  &    0.46   &    317.75  & -15.82   &   361.4   & 15.1   &   0.04  &  -1.82  \\
232.0021  & -50.6728  & NGC 5927  &              &      16.721   &    16.60  & 15.81  &    -5.02  &    0.15   &      -3.00  &    0.15   &    326.60  &   4.86   &  -115.7   &  7.4   &   0.47  &  -0.37  \\
246.8087  & -26.0247  & NGC 6144  &              &      33.352   &    16.60  & 16.01  &    -4.58  &    0.93   &     -11.68  &    2.35   &    351.93  &  15.70   &   188.9   & 10.1   &   0.32  &  -1.73  \\
272.3150  & -25.9078  & NGC 6553  &              &       8.135   &    16.60  & 15.79  &    -2.35  &    0.29   &      -2.35  &    0.27   &      5.25  &  -3.02   &    -6.5   &  4.7   &   0.78  &  -0.25  \\
211.3637  & +28.5344  & NGC 5466  &              &      52.754   &    16.62  & 16.10  &    -5.08  &    0.51   &      -2.10  &    0.55   &     42.15  &  73.59   &   107.7   & 16.6   &   0.00  &  -2.22  \\
199.1125  & +17.6981  & NGC 5053  &              &      14.866   &    16.65  & 16.14  &    -1.50  &    2.00   &      -2.10  &    1.85   &    335.69  &  78.94   &    44.0   & 16.2   &   0.03  &  -2.29  \\
249.8562  & -28.3978  & 1636-283  &  ESO452-SC11 &       7.697   &    16.66  & 15.96  &    -7.30  &    1.60   &      -3.50  &    1.40   &    351.91  &  12.10   &    17.5   &  7.6   &   0.50  &  -1.15  \\
\end{tabular}
\label{gctable}
\end{table*}

\addtocounter{table}{-1}
\begin{table*}[]
\scriptsize
\caption{End}
\label{Data2}
\medskip
\begin{tabular}{r r l l r r r r r r r r r r r r r}
  \hline
RA2000 & DEC2000 & Name &   Alternative &    $R_{tidal}$ &    V(HB) & DM &     pmra &   $e_{pmra}$ & pmde & $e_{pmde}$ &  l &  b &  $V_r$ &    Rh  &  $E_{B-V}$ & [Fe/H] \\
 & & & name & arcmin & & &  \multicolumn{4}{c}{mas/yr} & & & km/s& kpc & & \\
253.3558  & -22.1772  & NGC 6235  &              &       7.262   &    16.70  & 16.06  &    -3.52  &    1.26   &      -5.89  &    1.29   &    358.92  &  13.52   &    87.3   &  9.7   &   0.36  &  -1.40  \\
048.0637  & -55.2169  & NGC 1261  &              &       9.171   &    16.70  & 16.05  &     1.41  &    0.76   &      -2.67  &    0.71   &    270.54  & -52.13   &    68.2   & 16.0   &   0.01  &  -1.35  \\
308.5483  &  +7.4042  & NGC 6934  &              &      15.811   &    16.90  & 16.28  &    -1.92  &    0.32   &      -5.25  &    0.34   &     52.10  & -18.89   &  -411.4   & 15.2   &   0.12  &  -1.54  \\
198.2304  & +18.1692  & NGC 5024  &  M 53        &       8.471   &    16.90  & 16.36  &    -0.66  &    1.07   &      -0.30  &    0.80   &    332.96  &  79.76   &   -79.1   & 18.4   &   0.01  &  -2.07  \\
313.3662  & -12.5369  & NGC 6981  &  M 72        &      22.295   &    16.90  & 16.28  &    -0.85  &    0.49   &      -3.02  &    0.44   &     35.16  & -32.68   &  -345.1   & 16.8   &   0.05  &  -1.54  \\
260.2925  & -19.5872  & NGC 6342  &              &      10.351   &    16.90  & 16.15  &    -2.24  &    2.57   &      -8.23  &    2.54   &      4.90  &   9.73   &    80.9   &  9.1   &   0.44  &  -0.65  \\
256.2892  & -22.7081  & NGC 6287  &              &       6.310   &    17.00  & 16.46  &    -4.62  &    0.54   &      -1.65  &    0.63   &      0.13  &  11.02   &  -208.0   &  8.4   &   0.59  &  -2.05  \\
182.5258  & +18.5419  & NGC 4147  &              &      17.549   &    17.01  & 16.43  &    -0.63  &    2.15   &      -0.62  &    2.12   &    252.85  &  77.19   &   183.2   & 18.8   &   0.02  &  -1.83  \\
273.4121  & -31.8264  & NGC 6569  &              &      17.419   &    17.10  & 16.38  &    -1.78  &    0.28   &      -7.45  &    0.24   &      0.48  &  -6.68   &   -28.1   &  8.5   &   0.56  &  -0.86  \\
267.5537  & -37.0511  & NGC 6441  &              &       6.890   &    17.10  & 16.33  &    -4.07  &    0.30   &      -4.40  &    0.34   &    353.53  &  -5.01   &    18.3   &  9.7   &   0.45  &  -0.53  \\
326.6617  & -21.2508  & Pal 12    &              &      33.043   &    17.13  & 16.42  &    -0.77  &    0.73   &      -2.36  &    0.69   &     30.51  & -47.68   &    27.8   & 18.7   &   0.02  &  -0.93  \\
264.4004  &  -3.2458  & NGC 6402  &  M 14        &      15.811   &    17.20  & 16.56  &    -3.45  &    1.70   &      -6.57  &    1.50   &     21.32  &  14.81   &   -66.1   &  8.7   &   0.60  &  -1.39  \\
264.0708  & -44.7350  & NGC 6388  &              &      13.554   &    17.25  & 16.49  &    -1.00  &    0.16   &      -2.52  &    0.16   &    345.56  &  -6.74   &    81.2   & 11.5   &   0.38  &  -0.60  \\
280.3746  & -19.8258  & Pal 8     &              &      22.136   &    17.27  & 16.49  &    -3.13  &    0.27   &      -5.89  &    0.29   &     14.10  &  -6.80   &   -43.0   & 12.4   &   0.33  &  -0.48  \\
256.1200  & -24.7647  & NGC 6284  &              &       9.796   &    17.30  & 16.65  &    -3.00  &    0.28   &      -0.30  &    0.27   &    358.35  &   9.94   &    29.7   & 14.3   &   0.28  &  -1.32  \\
296.3100  &  -8.0072  & Pal 11    &              &       7.586   &    17.35  & 16.56  &    -0.94  &    0.55   &      -2.97  &    0.53   &     31.81  & -15.58   &   -68.0   & 12.6   &   0.34  &  -0.39  \\
301.5200  & -21.9214  & NGC 6864  &  M 75        &       7.975   &    17.47  & 16.82  &    -0.51  &    0.87   &      -2.17  &    0.87   &     20.30  & -25.75   &  -189.3   & 18.4   &   0.16  &  -1.32  \\
287.8004  &  +1.0306  & NGC 6760  &              &      15.937   &    17.50  & 16.73  &     0.87  &    0.31   &      -2.51  &    0.31   &     36.11  &  -3.92   &   -27.5   &  7.3   &   0.78  &  -0.52  \\
260.8958  & -17.8131  & NGC 6356  &              &      12.838   &    17.50  & 16.73  &    -4.40  &    0.17   &      -5.12  &    0.19   &      6.72  &  10.22   &    27.0   & 14.6   &   0.29  &  -0.50  \\
225.0771  & -82.2136  & IC 4499   &              &      12.244   &    17.65  & 17.04  &     0.12  &    0.79   &      -3.20  &    0.73   &    307.35  & -20.47   &    31.5   & 18.4   &   0.23  &  -1.60  \\
267.7158  & -34.5986  & NGC 6453  &              &       8.360   &    17.70  & 17.08  &    -0.96  &    0.24   &      -4.00  &    0.25   &    355.72  &  -3.87   &   -83.7   & 10.9   &   0.61  &  -1.53  \\
264.6537  & -23.9089  & NGC 6401  &              &      22.136   &    17.70  & 17.02  &    -4.18  &    0.26   &      -2.32  &    0.28   &      3.45  &   3.98   &   -65.0   &  7.5   &   0.85  &  -1.12  \\
217.4054  &  -5.9764  & NGC 5634  &              &       7.334   &    17.75  & 17.17  &    -5.25  &    2.20   &      -1.87  &    2.23   &    342.21  &  49.26   &   -45.1   & 25.3   &   0.05  &  -1.82  \\
289.4321  & -34.6575  & Terzan 7  &              &       6.032   &    17.76  & 17.00  &    -1.74  &    0.73   &      -2.70  &    0.65   &      3.39  & -20.07   &   166.0   & 23.0   &   0.06  &  -0.58  \\
259.1558  & -28.1400  & NGC 6316  &              &       5.012   &    17.78  & 17.01  &    -4.52  &    0.48   &      -3.70  &    0.47   &    357.18  &   5.76   &    71.5   & 11.5   &   0.55  &  -0.55  \\
189.6675  & -51.1503  & Rup 106   &              &      25.298   &    17.80  & 17.22  &    -5.31  &    0.53   &      -4.57  &    0.55   &    300.89  &  11.67   &   -44.0   & 20.6   &   0.21  &  -1.80  \\
233.8687  & -50.6594  & NGC 5946  &              &       2.194   &    17.80  & 17.16  &    -5.61  &    0.41   &      -1.38  &    0.43   &    327.58  &   4.19   &   119.5   & 12.3   &   0.55  &  -1.38  \\
246.9183  & -38.8489  & NGC 6139  &              &       1.328   &    18.00  & 17.40  &    -8.04  &    0.29   &      -2.93  &    0.29   &    342.37  &   6.94   &     6.7   & 10.5   &   0.74  &  -1.65  \\
251.7454  & +47.5278  & NGC 6229  &              &       8.833   &    18.00  & 17.37  &    -1.48  &    2.50   &       1.45  &    1.76   &     73.64  &  40.31   &  -154.2   & 29.3   &   0.01  &  -1.44  \\
295.4375  & -34.0003  & Terzan 8  &              &      13.031   &    18.03  & 17.46  &    -1.31  &    0.67   &      -3.25  &    0.67   &      5.76  & -24.56   &   130.0   & 25.4   &   0.14  &  -1.87  \\
283.7637  & -30.4783  & NGC 6715  &  M 54        &       6.325   &    18.17  & 17.56  &    -2.82  &    0.78   &      -2.47  &    0.34   &      5.61  & -14.09   &   141.9   & 26.2   &   0.15  &  -1.59  \\
254.8858  & -37.1214  & NGC 6256  &              &       9.487   &    18.20  & 17.45  &    -2.42  &    0.52   &      -0.97  &    0.48   &    347.79  &   3.31   &   -99.5   &  9.3   &   0.84  &  -0.70  \\
259.4967  & -23.7658  & NGC 6325  &              &      21.498   &    18.30  & 17.63  &    -8.26  &    1.24   &      -6.63  &    1.06   &      0.97  &   8.00   &     3.1   &  9.4   &   0.89  &  -1.17  \\
271.2075  &  -7.5858  & NGC 6539  &              &      21.058   &    18.33  & 17.58  &    -7.26  &    4.10   &      -6.96  &    2.96   &     20.80  &   6.78   &   -45.6   &  7.9   &   1.00  &  -0.66  \\
272.6842  &  -7.2075  & IC 1276   &  Pal 7       &      14.092   &    18.40  & 17.70  &    -6.46  &    0.54   &      -6.57  &    0.51   &     21.83  &   5.67   &   155.0   &  9.3   &   0.92  &  -0.56  \\
225.9937  & -33.0678  & NGC 5824  &              &       4.151   &    18.45  & 17.88  &    -1.98  &    1.80   &      -1.01  &    2.21   &    332.55  &  22.07   &   -38.4   & 31.3   &   0.13  &  -1.85  \\
219.9021  & -26.5383  & NGC 5694  &              &       9.487   &    18.50  & 17.93  &    -1.35  &    1.52   &      -1.43  &    1.29   &    331.06  &  30.36   &  -145.8   & 33.9   &   0.09  &  -1.86  \\
267.2192  & -20.3594  & NGC 6440  &              &       5.914   &    18.70  & 17.90  &    -3.96  &    1.28   &      -4.08  &    0.97   &      7.73  &   3.80   &   -78.7   &  8.0   &   1.09  &  -0.34  \\
247.1671  & -35.3536  & Terzan 3  &              &       6.313   &    18.80  & 18.10  &    -5.71  &    1.28   &      -2.66  &    0.77   &    345.08  &   9.19   &  -136.3   & 26.4   &   0.32  &  -0.73  \\
264.0437  & -38.5533  & Ton 2     &  Pismis 26   &       3.964   &    19.10  & 18.33  &    -1.64  &    0.89   &      -4.36  &    1.27   &    350.80  &  -3.42   &  -184.4   &  7.9   &   1.24  &  -0.50  \\
136.9908  & -37.2214  & Pyxis     &              &      10.436   &    19.25  & 18.58  &    -2.00  &    0.90   &       2.80  &    1.30   &    261.32  &   7.00   &    34.3   & 38.5   &   0.21  &  -1.20  \\
270.4546  & -27.8258  & Djorg 2   &  E456-SC38   &       5.206   &    19.50  & 18.80  &    -3.26  &    0.29   &      -3.47  &    0.28   &      2.76  &  -2.51   &  -150.0   & 13.8   &   1.00  &  -0.50  \\
286.3137  &  +1.9008  & NGC 6749  &              &       8.309   &    19.70  & 19.09  &    -3.80  &    0.39   &      -4.60  &    0.35   &     36.20  &  -2.20   &   -61.7   &  7.7   &   1.50  &  -1.60  \\
265.9258  & -26.2225  & Pal 6     &              &       3.080   &    19.70  & 18.87  &    -7.72  &    0.46   &      -6.96  &    0.47   &      2.09  &   1.78   &   201.0   &  6.7   &   1.53  &  -0.10  \\
261.7854  &  -7.0931  & IC 1257   &              &      12.649   &    19.80  & 19.20  &    -3.54  &    0.64   &      -1.69  &    0.63   &     16.53  &  15.14   &   140.2   & 24.5   &   0.73  &  -1.70  \\
263.9492  & -30.4697  & Terzan 1  &  HP 2        &       4.817   &    19.95  & 19.15  &     1.44  &    0.75   &      -3.47  &    0.41   &    357.57  &   1.00   &    35.0   &  6.5   &   1.64  &  -0.35  \\
261.8892  & -30.8022  & Terzan 2  &  HP 3        &       3.147   &    20.10  & 19.29  &    -3.38  &    0.28   &      -3.63  &    0.29   &    356.32  &   2.30   &   109.0   &  9.5   &   1.42  &  -0.25  \\
266.8679  & -33.0656  & Djorg 1   &              &       2.038   &    20.80  & 20.10  &    -7.97  &    0.63   &      -3.73  &    0.60   &    356.67  &  -2.48   &  -362.4   &  9.2   &   1.70  &  -0.40  \\
071.5246  & +31.3808  & Pal 2     &              &       9.892   &    21.65  & 20.99  &     7.19  &    0.62   &       0.27  &    0.63   &    170.53  &  -9.07   &  -133.0   & 26.9   &   1.24  &  -1.30  \\
268.6133  & -24.1453  & UKS 1     &              &      11.972   &    24.14  & 23.47  &    -2.78  &    0.51   &      -0.55  &    0.54   &      5.12  &   0.76   &    57.0   &  7.5   &   2.93  &  -1.20  \\
\end{tabular}
\end{table*}

\section{Galactic potential model}
\label{Model}
In this paper we use a model gravitational potential of the Galaxy, which includes both axisymmetric and non-axisymmetric 
parts. The axisymmetric part is represented by three components: the Miyamoto and Nagai disk \citep{Miyamoto}, 
Hernquist spheroid \citep{Hernquist}, and modified isothermal dark-matter halo. The formulas for the potentials of these three
components have the following form:

\begin{center}
	\begin{equation}
		\phi_{disk} = -\frac{GM_{disk}}{\sqrt{R^2+{\left( a+\sqrt{z^2+b^2} \right)}^2}}
	\end{equation}
	\begin{equation}
		\phi_{spher} = -\frac{GM_{spher}}{\sqrt{R^2+z^2}+c}
	\end{equation}
	\begin{equation}
		\phi_{halo} = V_{halo}^2\ln\left(R^2+z^2+d^2\right)
	\end{equation}
\end{center}

Here we use the following parameter values: $M_{disk}=10^{11}M_{sun}, a=5, b=0.26, M_{spher}=3.4\times10^{10}M_{sun}, 
c=0.7, V_{halo}=1.15, d=12$ (all distances are in kpc). We choose these constants so as to ensure that the resulting rotation curve in
the Galactic disk would fit the rotation curve based on recent data about the kinematics of Galactic masers~\citep{Rastorguev-Utkin}, 
implying a circular velocity rotation of about 237~km/s at the solar circle
(we adopt $R_0=8.3$~kpc throughout this paper).

\begin{figure}
	\includegraphics[width=1.1\linewidth]{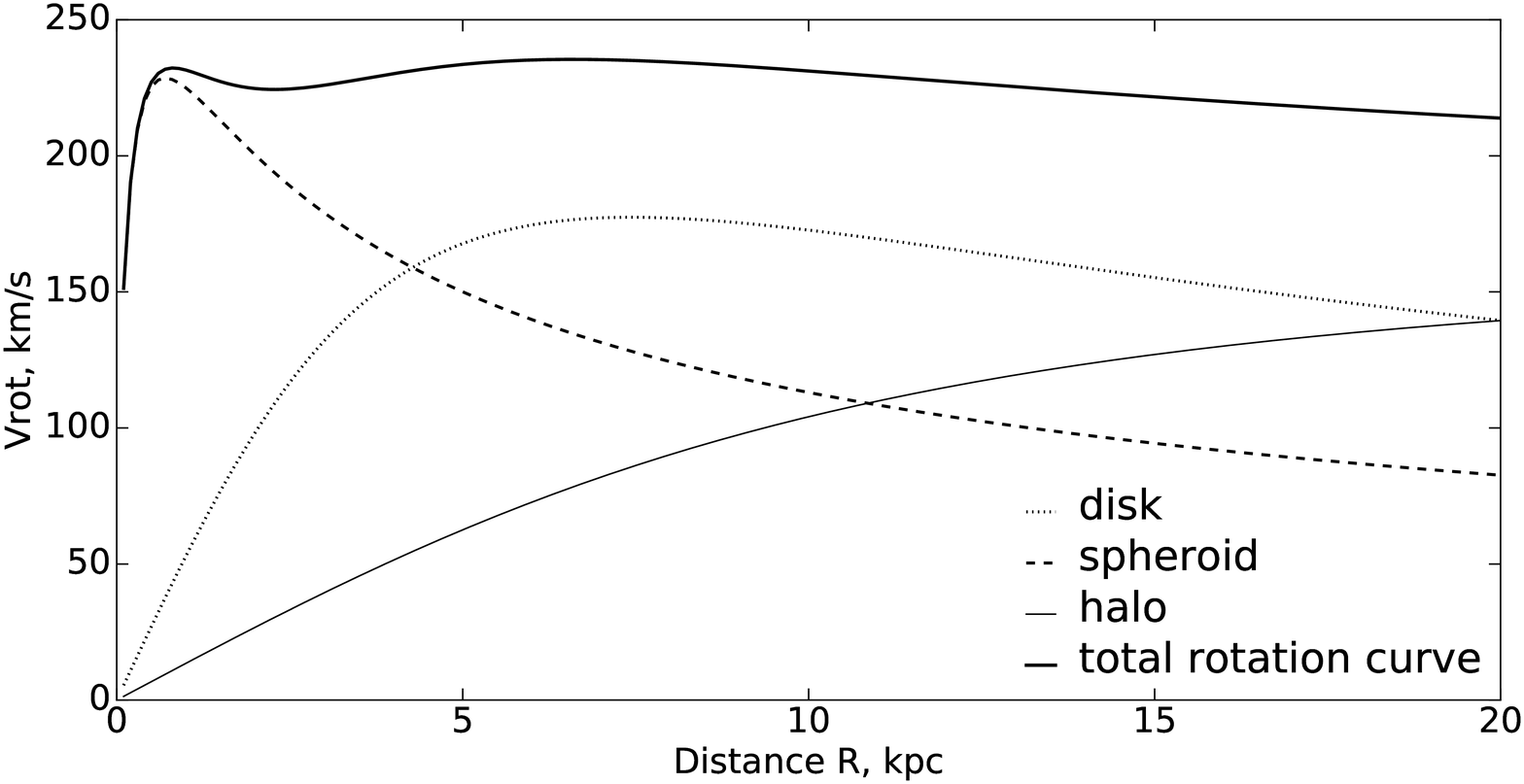}
	\caption{Decomposition of the rotation curve based on the axisymmetric part of the Galactic gravitational potential}
	\label{rot_curve}
\end{figure}

The non-axisymmetric part of the gravitational potential is represented by the Galactic bar, which we modeled by a Ferrer's bar 
with index $n=2$ \citep{Vaucouleurs-Freeman}. In this model the bar has the shape of an ellipsoid of rotation 
with the semimajor and semiminor axes equal to $a_{bar}=4$~kpc and $c_{bar}=1$~kpc, respectively, and the axis of rotation
located in the symmetry plane of the Galaxy. The density distribution in the adopted bar model has the form:

\begin{center}
	\begin{equation}
		\rho=
		\begin{cases}
			\rho_0{\left( 1 - m^2 \right)}^2 & \text{$m < 1$}\\
			0 & \text{$m \geq 1$}
		\end{cases}
	\end{equation}
\end{center}

where $m^2=x^2/a_{bar}^2 + \left( y^2+z^2 \right)/c_{bar}^2$. The formula for the gravitational potential in the Cartesian
coordinate system associated with the bar (the $x$-axis  directed along the semimajor axis of the bar,
the $z$-axis points toward the North Galactic Pole, and the $x$-axis  directed along the semiminor axis of the bar  
so as to make the right-handed coordinate system) has the following form:
\begin{align}
\label{eq1}
	\phi_{bar} &= -\frac{105GM_{bar}}{32\varepsilon}\Big[  \frac{1}{3}w_{10}-\Big( \Big( y^2+z^2 \Big)w_{20}+\nonumber\\ & +x^2w_{11} \Big)+
	\Big( {\Big( y^2+z^2 \Big)}^2w_{30}+2\Big( y^2+z^2 \Big)x^2w_{21}+\nonumber\\ & +x^4w_{12} \Big) - \frac{1}{3}\Big( {\Big( y^2+z^2 \Big)}^3w_{40} + 3{\Big( y^2+z^2 \Big)}^2x^2w_{31}+\nonumber\\ & + 3\Big( y^2+z^2 \Big)x^4w_{22} + x^6w_{13}\Big) \Big]
\end{align}

where $\varepsilon^2=a_{bar}^2-c_{bar}^2$. The coordinates $x, y, z$ in equation~\ref{eq1} are nondimensionalized
by dividing them by parameter $\varepsilon$.
The integrated coefficients  $w_{ij}$ are defined as:
\begin{center}
	\begin{equation}
		w_{ij}\left(\psi\right)=2\int\limits_0^\psi \tan^{2i-1}\theta \sin^{2j-1}\theta \mathrm{d}\theta
	\end{equation}
\end{center}
where $\psi\left(x,y,z\right)$ is the function of coordinates equal to the solution of the equation
\begin{align}
	\left(y^2+z^2\right)\tan^2\psi+x^2\sin^2\psi &=\varepsilon^2 &\text{$m > 1$}\nonumber\\
	\cos\psi &= c_{bar}/a_{bar} &\text{$m \leq 1$}
\end{align}
Throughout this study we assumed that at the present time the orientation angle of the bar with respect to the
Galactic center--Sun direction is equal to 45\degr  (the first quadrant).
We assumed the angular velocity and mass of the bar to be 50 km/s/kpc and  10\% of the Galactic disk mass
($M_{bar}=10^{10}M_{sun}$). 

\section{Orbits}
\label{Orbits}
\begin{figure*}[]
\includegraphics[width=\linewidth]{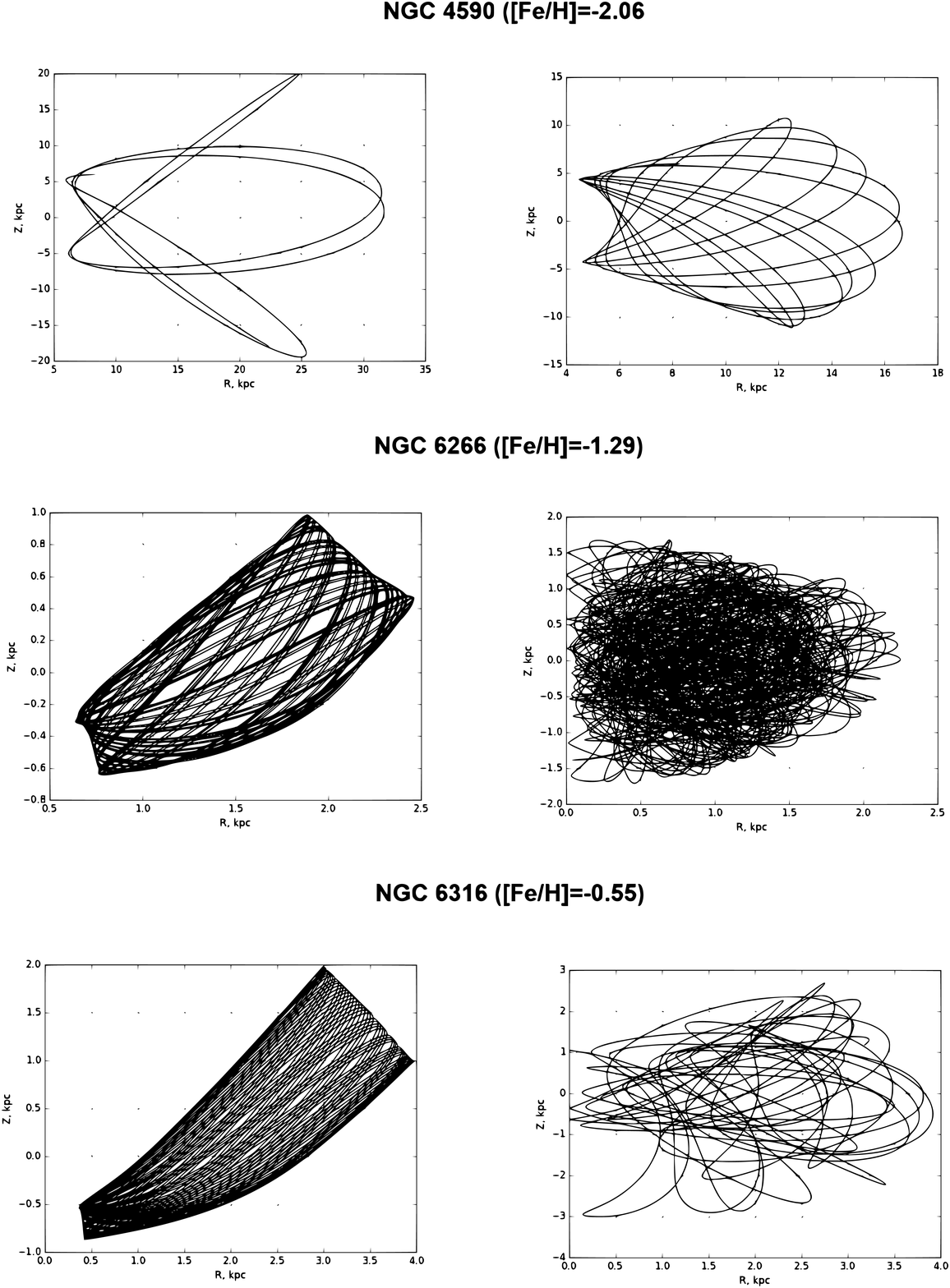}
\caption{Meridional cross sections of the orbits of the globular clusters NGC~4590, NGC~6266, and NGC~6316) computed with
axisymmetric potential (on the left) and with the barred potential (on the right).
Chaotization caused by the  bar is especially evident in the case of NGC~6266 and NGC~6316.} \label{fig:meridional}
\end{figure*}

\begin{figure*}[]
\includegraphics[width=0.8\linewidth]{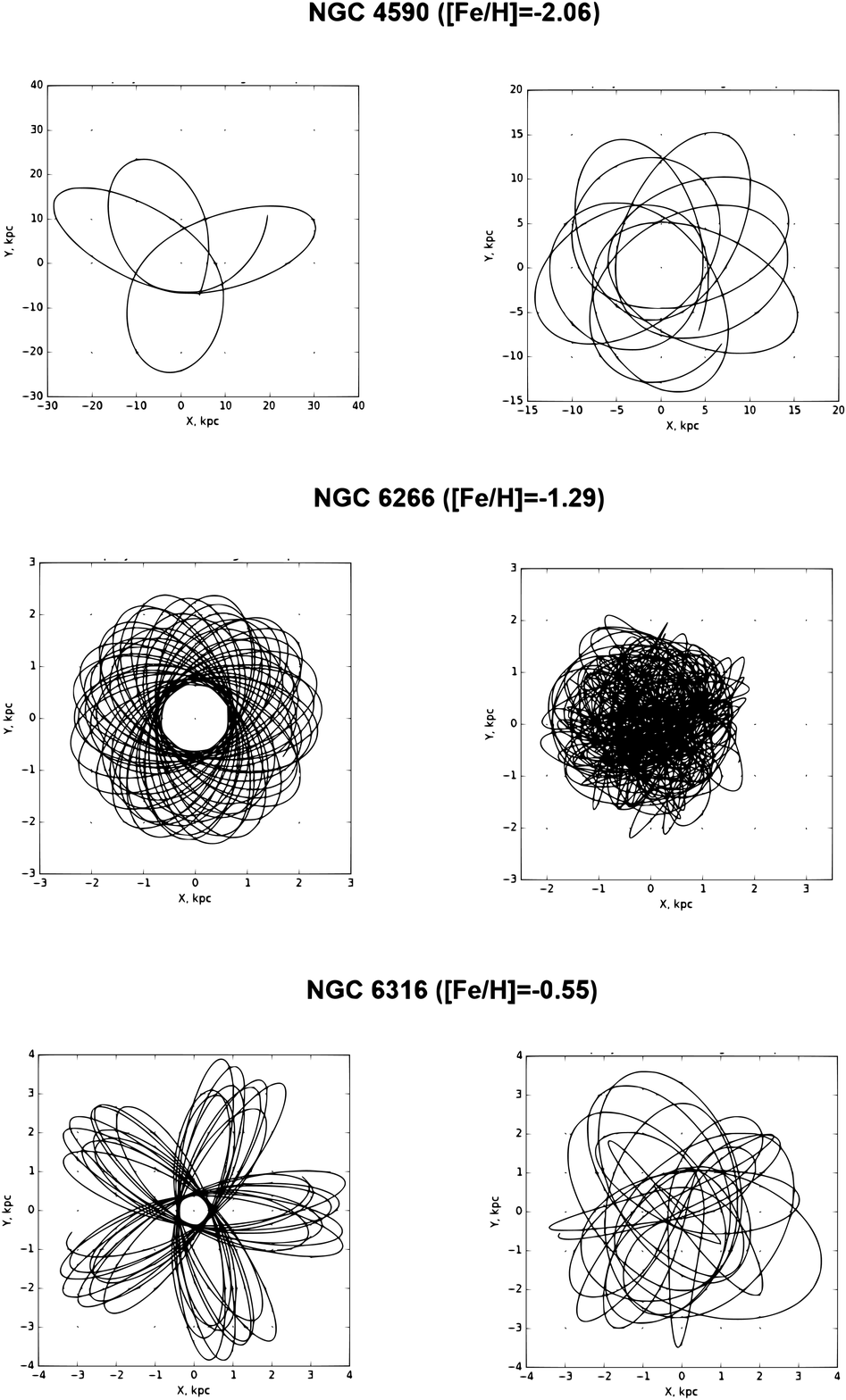}
\caption{Galactic-plane projections of the orbits of the globular clusters NGC~4590, NGC~6266, and NGC~6316) computed with
axisymmetric potential (on the left) and with the barred potential (on the right).
Like in the case of meridional cross sections,
chaotization caused by the  bar is especially evident in the case of NGC~6266 and NGC~6316.} \label{fig:galplane}
\end{figure*}

\begin{figure}[]
\includegraphics[width=\linewidth]{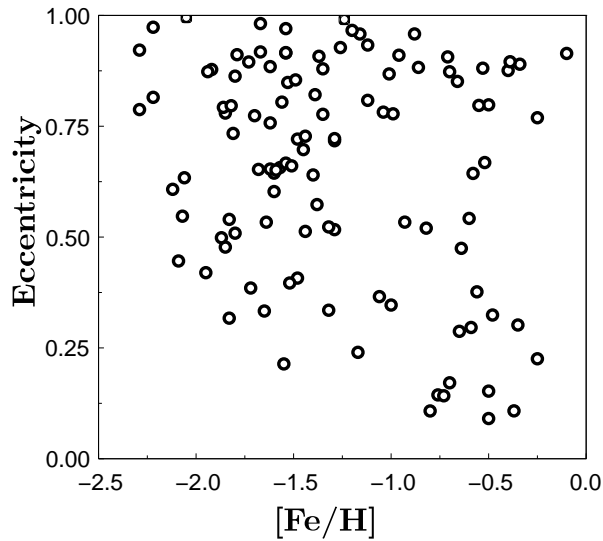}
\caption{Metallicity vs. average eccentrcity plot for orbits computed in terms of axisymmetric potential model.} \label{fig:exmetal1}
\end{figure}

\begin{figure}[]
\includegraphics[width=\linewidth]{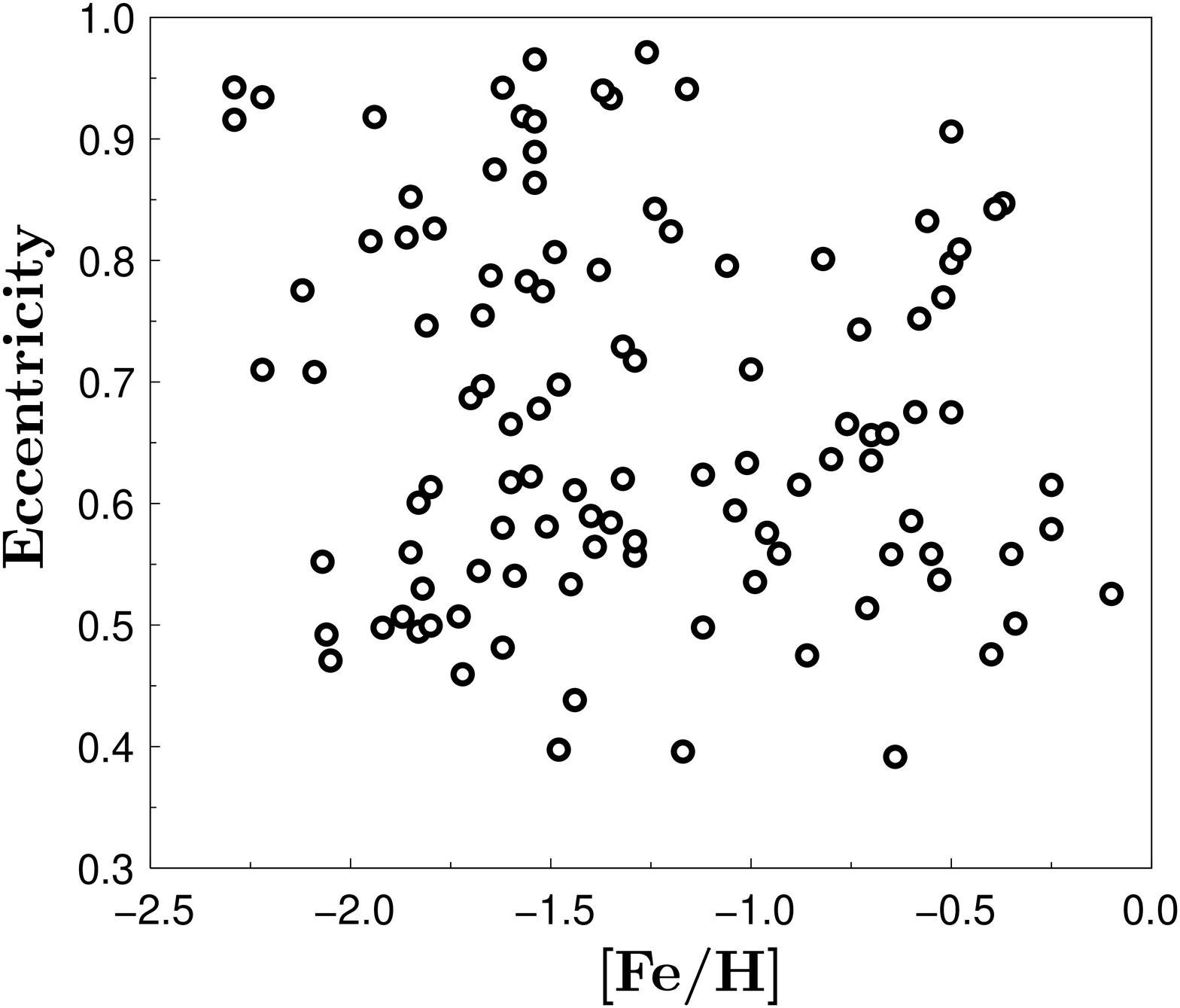}
\caption{Metallicity vs. average eccentrcity plot for orbits computed in terms of barred potential model.} \label{fig:exmetal2}
\end{figure}

We computed the orbits of 115 Milky-Way globular clusters in terms of two models of the gravitational potential
of the Galaxy: a purely axisymmetric model (disk+spheroid+halo) and a model, which, in addition to the above three
components includes a rotating bar. Five of 115 clusters --  Terzan 3, NGC 5634, Rup 106, Pyxis, and Pal 2 -- have escaping
trajectories, which are most likely due to to large errors of the inferred proper motions of these clusters resulting
in the total velocities greater than the escape velocities at the corresponding locations (all these clusters are 
quite far away from the Sun at heliocentric distances ranging from 20.6 to 38.5~kpc, which explain the large errors of their
proper motions and transversal velocities). We computed the orbits of all clusters orbits for 2-Gyr forward, except for
IC 1257, NGC 6101, NGC 6229, and NGC 6715. The latter four clusters are too distant to make more than one revolution within 2~Gyr 
and we integrated their orbits for 5~Gyr forward. The parameters of the resulting globular-cluster orbits are listed
in Tables~\ref{Res1} (for the model potential without the bar) and \ref{Res2} (for the model potential with a bar). The full versions of
Tables~\ref{Res1} and ~\ref{Res2} and images of the orbits of all clusters are available in electronic form at
\newline
\verb|www.sai.msu.ru/groups/cluster/cl/orbits_gcl/|
\newline 
Table~\ref{Res1} gives the following quantities: cluster name;  ${\left( R_{min} \right)}_{min}$ and $<R_{min}>$
(the minimum and average pericentric distance, respectively);${\left( R_{max} \right)}_{max}$ and $<R_{max}>$ 
(the maximum and average apocentric distance, respectively); 
${\left( |z|_{max} \right)}_{max}$ and $<|z|_{max}>$ (the maximum and average distance from the symmetry plane of the
Galaxy, respectively); $<e>$, the estimated average eccentricity of the cluster orbit; $E$, the total mechanical
energy per unit mass of the cluster in ${100 \text{km/s}}^2$, and $h$, the projection of the specific angular momentum
of the cluster onto the symmetry axis of the Galaxy (in 100 kpc~$\cdot$~km/s). All distances are in kpc. 
In the case of the axisymmetric model potential  $E$ and $h$ are conserved. Table~\ref{Res2} differs from Table~\ref{Res2} in that
it gives the minimum, maximum, and average values of  $E$ and $h$  ($E_{min}$, $E_{max}$, $E_{avg}$,
$h_{min}$, $h_{max}$, and $h_{avg}$) because the total mechanical energy and projection of the orbital momentum onto the
Galactic symmetry axis are nor conserved in the case of the barred potential (the bar brings explicit dependence on time and angle into
the Hamiltonian of the cluster). When computing the total mechanical energy of the cluster we set  the
gravitational potential equal to zero at the Galactocentric distance of 300~kpc (adopted boundary of the Galaxy).
Figs.~\ref{fig:meridional} and \ref{fig:galplane} show  the meridional cross sections and galactic-plane projections 
of three globular clusters (the metal-poor clusters NGC~4590 and NGC~6266 and the metal-rich cluster
NGC~6316) computed with axisymmetric potential (the figures on the left) and with the barred potential (the figures
on the right). Interestingly, the correlation between the metallicity and average eccentricity, which shows up
conspicuously in for orbits computed with axisymmetric potential with the most metal-poor clusters have almost exclusively
high eccentricties (Fig.~\ref{fig:exmetal1}) disappears if the orbits are computed in the barred potential (Fig.~\ref{fig:exmetal2}).

\begin{figure*}[]
\includegraphics[width=\linewidth]{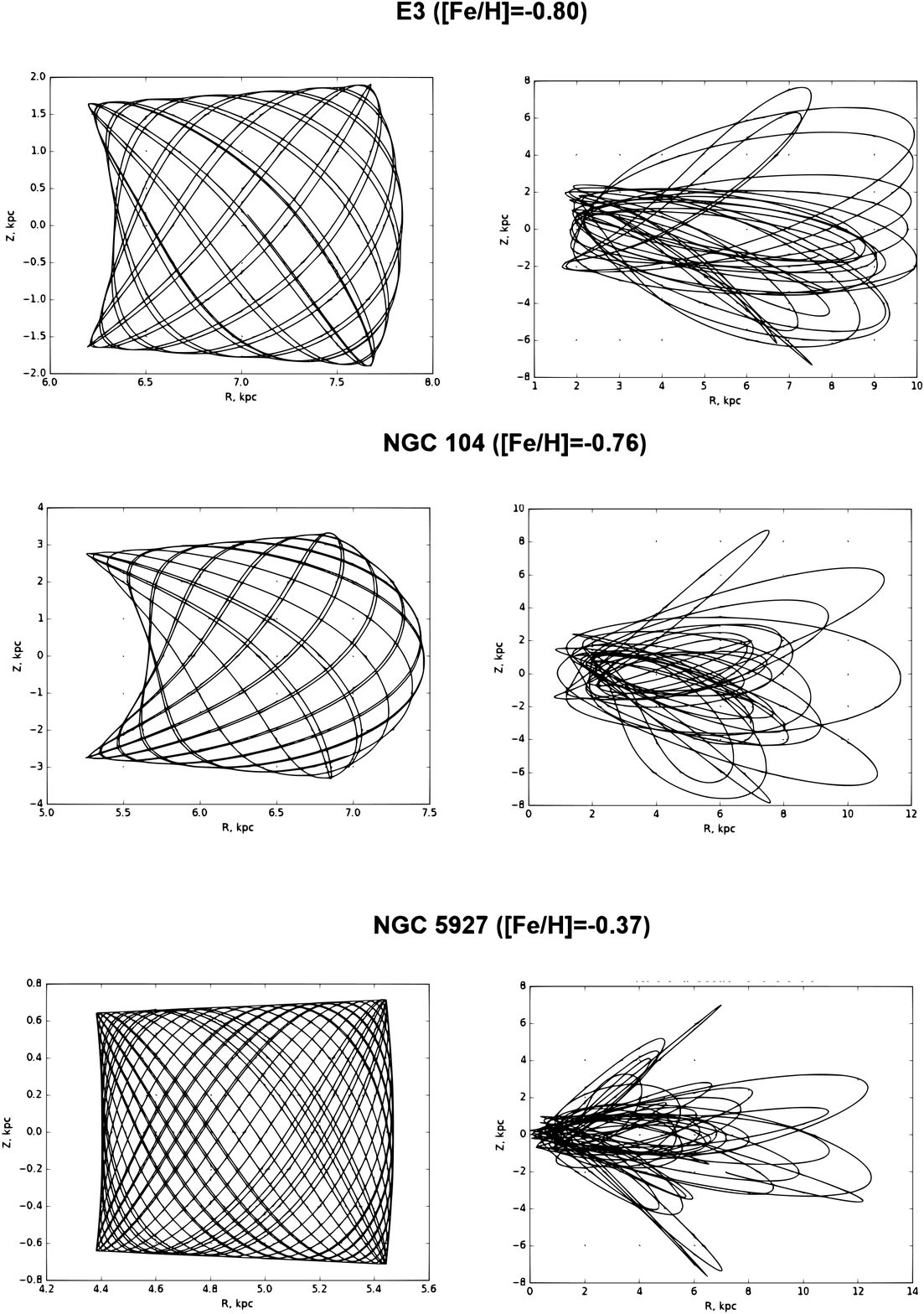}
\caption{Meridional cross sections of the orbits of the globular clusters E3, NGC~104, and NGC~5927) computed with
axisymmetric potential (on the left) and with the barred potential (on the right).} \label{fig:meridional2}
\end{figure*}

\begin{figure*}[]
\includegraphics[width=0.8\linewidth]{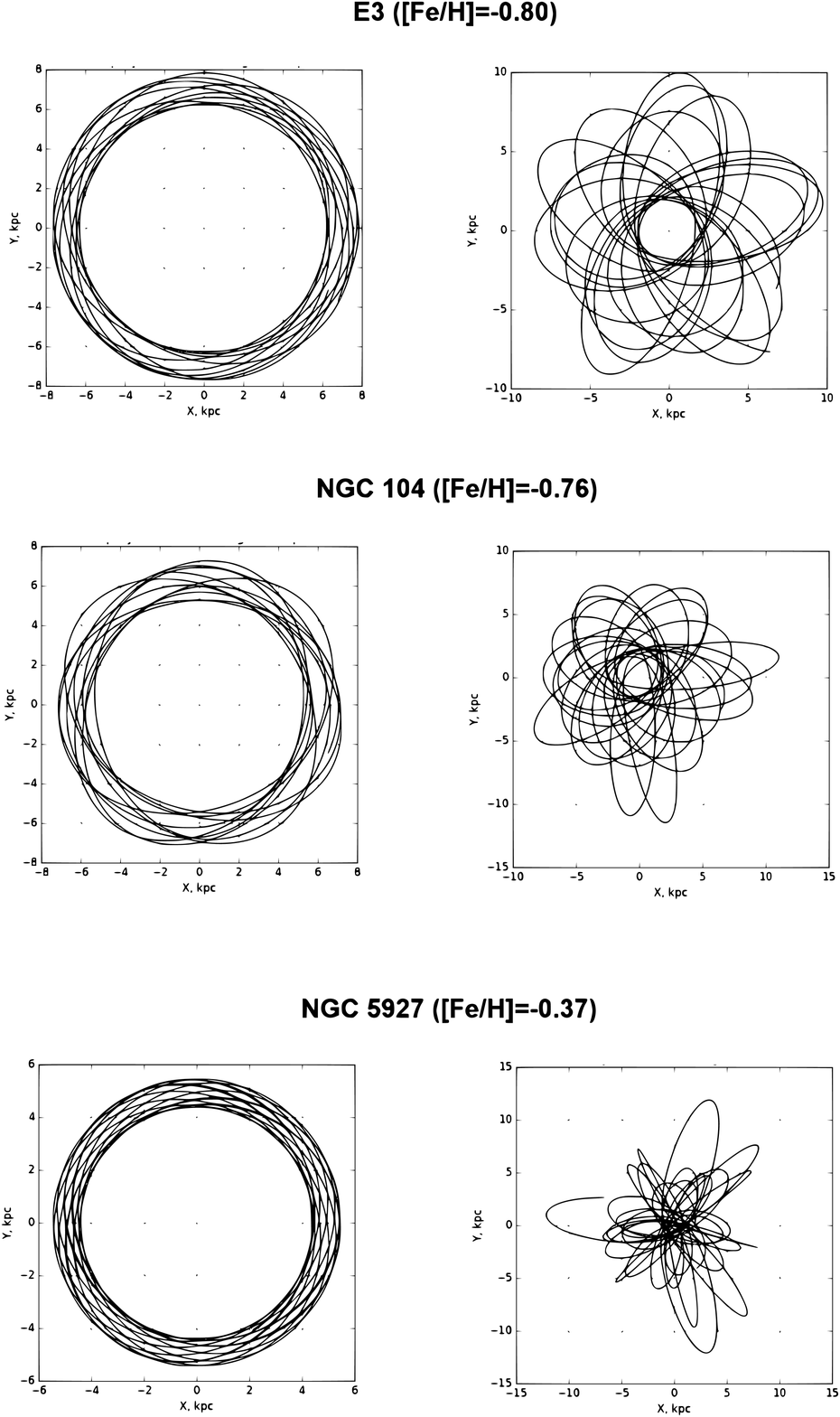}
\caption{Galactic-plane projections of the orbits of the globular clusters E3, NGC~104, and NGC~5927) computed with
axisymmetric potential (on the left) and with the barred potential (on the right).} \label{fig:galplane2}
\end{figure*}

Unlike the conclusions reached by the authors of the so far most extensive observational study of the orbits of a total of 54
Galactic globular clusters~\citep{Allen1, Allen2}, we find that the bar has appreciable effect on the orbits of all clusters: it
destroys the orbital boxes and randomizes (chaotizes) the orbits. The effect of the bar on distant orbits
is weaker (central parts of the orbits mostly get entangled), which is to be expected given the 10\% contribution
of the bar to the effective mass. Orbits of some of the thick-disk clusters are literally "stretched out" by the bar and
become, on the average, closer to the Galactic center. Typical examples include such clusters as E3, NGC~104, and NGC~5927 whose
orbits we show in Figs.~\ref{fig:meridional2} and \ref{fig:galplane2}.
We defer a more detailed analysis of the cluster orbits to our forthcoming paper.

\begin{table*}[]
\scriptsize
\caption{Results of the analysis of cluster orbits in the potential model without the bar}
\label{Res1}
\medskip
\begin{tabular}{l r r r r r r r r r}
  \hline
Name & ${\left( R_{min} \right)}_{min}$ & $<R_{min}>$ & ${\left( R_{max} \right)}_{max}$ & $<R_{max}>$ & ${\left( |z|_{max} \right)}_{max}$ & $<|z|_{max}>$ & 
$<e>$ & $E$ & $h$ \\
1636-283 & 0.04 & 0.05 & 3.10 & 2.43 & 2.11 & 0.88 & 0.96 & -18.89 & 0.22 \\
IC 1257 & 7.15 & 8.57 & 86.05 & 67.22 & 70.86 & 44.74 & 0.77 & -3.56 & 26.19 \\
IC 4499 & 4.28 & 7.62 & 61.37 & 35.20 & 58.84 & 33.44 & 0.64 & -4.53 & 13.18 \\
\end{tabular}
\label{res_no_bar}
\end{table*}

\begin{table*}[]
\scriptsize
\caption{Results of the analysis of cluster orbits in the potential model with a bar}
\label{Res2}
\medskip
\begin{tabular}{l r r r r r r r r r r r r r}
  \hline
Name & ${\left( R_{min} \right)}_{min}$ & $<R_{min}>$ & ${\left( R_{max} \right)}_{max}$ & $<R_{max}>$ & ${\left( |z|_{max} \right)}_{max}$ & $<|z|_{max}>$ & 
$<e>$ & $E_{min}$ & $E_{max}$ & $E_{avg}$ & $h_{min}$ & $h_{max}$ & $h_{avg}$ \\
NGC 1261 & 0.03 & 0.54 & 22.65 & 15.83 & 9.81 & 3.26 & 0.93 & -15.83 & -9.34 & -12.77 & -1.58 & 11.11 & 4.40 \\
NGC 1851 & 0.07 & 0.23 & 21.68 & 15.98 & 23.19 & 8.55 & 0.97 & -11.72 & -6.82 & -10.11 & -4.40 & 5.19 & -1.25 \\
NGC 1904 & 0.10 & 0.28 & 31.66 & 16.15 & 19.93 & 6.90 & 0.97 & -14.36 & -8.19 & -11.34 & -6.62 & 5.44 & -0.71 \\
\end{tabular}
\label{res_bar}
\end{table*}

\section{Conclusions}
\label{Conclusions}
We determined accurate absolute proper motions (with a typical accuracy of $\sim$~0.4~mas/yr,
which translates into a $\sim$~17~km/s transversal-velocity error) and computed Galactic orbits 
for the currently largest sample of Galactic globular clusters (115 objects), which represents
a two-fold increase compared to the most extensive previous studies. We computed the 
cluster orbits in terns of both an axisymmetric potential model and a model with
a rotating bar. Unlike what was found by the authors of earlier studies, we conclude that
the bar has appreciable effect on the orbits of practically all clusters in that
it randomizes the orbits and especially their portions in the vicinity of the Galactic center,
and stretches out the orbits of some of the thick-disk clusters.

\section*{ACKNOWLEDGEMENTS}

This work has made use of data from the European Space Agency (ESA)
mission {\it Gaia} (\url{https://www.cosmos.esa.int/gaia}), processed by
the {\it Gaia} Data Processing and Analysis Consortium (DPAC,
\url{https://www.cosmos.esa.int/web/gaia/dpac/consortium}). Funding
for the DPAC has been provided by national institutions, in particular
the institutions participating in the {\it Gaia} Multilateral Agreement.
This publication makes use of data products from the Two Micron 
All Sky Survey, which is a joint project of the University of Massachusetts 
and the Infrared Processing and Analysis Center/California Institute of Technology, 
funded by the National Aeronautics and Space Administration and the National Science Foundation,
and of the data products from the Wide-field Infrared Survey Explorer, which is a joint project 
of the University of California, Los Angeles, and the Jet Propulsion Laboratory/California 
Institute of Technology, and NEOWISE, which is a project of the Jet Propulsion Laboratory/California 
Institute of Technology. WISE and NEOWISE are funded by the National Aeronautics and Space Administration..
This work was supported by the Russian Foundation for Basic Research (grant no.~18-02-00890).




\clearpage

\onecolumngrid \clearpage

\end{document}